\documentclass[useAMS,usenatbib]{mn2e}
\usepackage{epsfig}

\def\msun{{\rm {M}}_{\odot}}
\newcommand{\etal}{{et al.}~}
\newcommand{\eg}{{e.g.~}}
\newcommand{\ie}{{i.e.~}}

\def \ltsima{$\; \buildrel < \over \sim \;$}
\def \simlt{\lower.5ex\hbox{\ltsima}}            
\def \gtsima{$\; \buildrel > \over \sim \;$}
\def \gtsima{\mbox{$\; \buildrel > \over \sim \;$}}
\def \simgt{\lower.5ex\hbox{\gtsima}}            

\newcounter{cureqno}

\title[clustering of the 1$^{st}$ galaxy halos]{
The clustering of the first galaxy halos\thanks{LA-UR 07-8350}}
\author[Reed \etal] {
Darren S. Reed$^{1}$ \thanks{Email:
reed@lanl.gov},
Richard Bower$^{2}$, 
Carlos S. Frenk$^{2}$,
\newauthor
~Adrian Jenkins$^{2}$, 
and Tom Theuns$^{2,3}$\\
$^1$Theoretical Astrophysics Group/ISR-1, Los Alamos National Laboratory,
PO Box 1663, MS 627, Los Alamos, NM, 87545 USA\\
$^2$Institute for Computational Cosmology, Dept. of Physics, 
University of Durham,  South Road, Durham DH1 3LE, UK\\ 
$^3$Dept. of Physics, Univ. of Antwerp, Campus Groenenborger, 
Groenenborgerlaan 171, B-2020 Antwerp, Belgium}

\pagerange{\pageref{firstpage}--\pageref{lastpage}}
\pubyear{xxxx}

\begin{document}

\maketitle

\label{firstpage}

\begin{abstract}

We explore the clustering properties of high redshift dark matter
haloes, focusing on haloes massive enough to host early generations of
stars or galaxies at redshift 10 and greater.  Haloes are extracted
from an array of dark matter simulations able to resolve down to the
``mini-halo'' mass scale at redshifts as high as 30, thus encompassing
the expected full mass range of haloes capable of hosting luminous
objects and sources of reionization.  Halo clustering on large-scales
agrees with the Sheth, Mo \& Tormen halo bias relation within all our
simulations, greatly extending the regime where large-scale
clustering is confirmed to be ``universal'' at the $10-20\%$ level
(which means, for example, that $3\sigma$ haloes of cluster mass at
$z=0$ have the same large-scale bias with respect to the mass
distribution as $3\sigma$ haloes of galaxy mass at $z=10$).  However,
on small-scales, the clustering of our massive haloes ($\simgt 10^{9}
h^{-1}\msun$) at these high redshifts is stronger than expected from
comparisons with small-scale halo clustering extrapolated from lower
redshifts.  This implies ``non-universality'' in the scale-dependence
of halo clustering, at least for the commonly used parameterizations
of the scale-dependence of bias that we consider.  We provide a fit
for the scale-dependence of bias in our results.  This study provides
a basis for using extraordinarily high redshift galaxies (redshift
$\sim$ 10) as a probe of cosmology and galaxy formation at its
earliest stages.  We show also that mass and halo kinematics are
strongly affected by finite simulation volumes.  This suggests the
potential for adverse affects on gas dynamics in hydrodynamic
simulations of limited volumes, such as is typical in simulations of
the formation of the ``first stars'', though further study is
warranted.

\end{abstract}

\section{introduction}

Dark matter haloes are formed from fluctuations in the matter density
field as characterized by the matter power spectrum.  A theoretical
understanding of the relationship between the matter power spectrum
and the numbers and clustering of dark matter haloes provides an
essential link between cosmological parameters and the properties of
haloes in which observable (or potentially observable) galaxies live.
Haloes form from gravitationally induced non-linear collapse of mass
overdensities that grow from primordial density fluctuations.  Thus,
halo clustering properties in a universe composed of cold dark matter
are determined entirely by the matter power spectrum, or equivalently,
by the cosmological parameters.  The aim of this paper is to use dark
matter simulations to quantify the relation between the matter power
spectrum and resulting halo clustering properties, focusing on haloes
of masses capable of star or galaxy formation during the era of
reionization.  It is beyond the scope of this paper to incorporate
baryon physics to model galaxy formation within haloes.  However, the
distribution of haloes, which we explore, can be used as a basis for
studies that populate haloes with galaxies to model galaxy clustering
properties, which can then be used to probe cosmology and galaxy
formation at high redshifts.

Halo clustering and its evolution has been studied extensively by a
number of authors, mainly focusing on cluster, group, or galaxy mass
scales at low redshifts (\eg \citealt{whiteclust}; \citealt{mowhite};
\citealt{jing}; \citealt{st99}; \citealt{colberg}; \citealt{taruya} ;
\citealt{yoshbias}; \citealt{hamana} ; \citealt{seljakbias};
\citealt{tinkerbias}; \citealt{wechslerbias}; \citealt{wetzelbias};
\citealt{angulobias}
\citealt{basbias}).  However, few studies have explored the halo or
galaxy distribution in the reionizing era (\eg \citealt{zahnreion};  
\citealt{cohnhiz};
\citealt{ilievclust}; \citealt{ricotti1st}), an epoch just beginning
to be targeted by a number of surveys, expected to provide a new probe
of cosmology and the physics of galaxy formation.  At these high
redshifts, haloes large enough to host stars and galaxies are expected
to form from much rarer fluctuations (\ie collapsed from higher sigma
overdensities) than haloes that host typical galaxies or even groups
and clusters in the low redshift universe.  This could lead to
different halo clustering properties, which we will explore using
numerical simulations.

The Press \& Schechter formalism (\citealt{ps}) provides a general
analytic framework for understanding halo formation, and has been
improved by a number of subsequent authors (\eg \citealt{bower};
\citealt{lc93}).  Utilizing the extended Press \& Schechter formalism,
\cite{mowhite} developed a theoretical means of predicting halo
clustering properties from the linear matter power spectrum.
Numerical simulations have been utilized by a number of the
fore-mentioned authors to validate and improve upon these works.
Recently, we used a suite of high resolution numerical simulations to
quantify the numbers of haloes during the reionizing era
\cite{reedmf07} (see also \eg Luki{\'c} \etal 2007 and
\citealt{heitmann}).  This work utilizes the same simulations to
analyze the clustering properties of haloes.

We argue that halo clustering is at least somewhat ``universal'' with
respect to the linear matter fluctuation field.  If halo clustering is
exactly ``universal'', then haloes of different mass and redshift
formed from linear fluctuations of the same rarity will have identical
clustering properties (see \S~2).  For example, $3\sigma$ haloes of
cluster mass at $z=0$ would have the same relative clustering with
respect to the mass distribution as $3\sigma$ haloes of galaxy mass at
$z=10$.  Universality of halo clustering has been explored at lower
redshifts by authors mentioned earlier, but has not been thoroughly
addressed during the reionization period.  We argue later that
approximate universality holds for large-scale halo clustering to redshift
30, though it breaks down at small scales for the more massive haloes
at z$\simgt$10.  The universality of large-scale clustering that we
will show is a convenient feature (also held by the halo mass
function; see \eg \citealt{jenkmf}; Reed \etal 2003, 2007;
Luki{\'c} \etal 2007) that should enable the results of our simulations,
which are evolved only to $z=10$, to be applicable to rare haloes over
a wide redshift range.  Rare, massive haloes attainable in current and
future generations of high redshift clustering surveys, should have
large-scale clustering strength relative to the mass, \ie {\it
bias}, similar to that of equally rare, massive haloes at lower
redshifts (\eg clusters, QSO hosts)

The mass range of high redshift haloes that we are able to resolve
includes virtually all haloes large enough to form stars or galaxies;
see {\it e.g.} \cite{reedfirststar} and references therein for a
review of high redshift star formation.  The first stars are expected
to form within ``mini-haloes'' of $\sim 10^{5-6} \msun$ where
primordial gas cools and contracts mainly by $H_{2}$-cooling (\eg
Abel, Bryan, \& Norman 2000,2002; Bromm, Coppi, \& Larson 1999,2002;
\citealt{yoshida1st}; \citealt{osheanorman}).  The ``first galaxies''
may form within haloes of virial temperatures greater than $\sim 10^4K$
($\sim 10^8 \msun$ at $z\sim10$), where gas is able to cool via
collisionally-induced atomic hydrogen cooling, a more effective
process.  One should keep in mind that metal enrichment or other
feedback effects may have a large influence on what types of haloes can
cool and transform gas into stars sufficiently to resemble a
``galaxy''.

Using extremely high redshift galaxies for cosmology has some
particular advantages.  Their small masses allow the matter power
spectrum to be probed on very small scales.  For example, warm dark
matter models with filtering lengths $\simlt 100 h^{-1}$kpc can be
tested if the abundance and bias of $10^8 h^{-1}\msun$ hosts of the
``1st galaxies'' can be measured.  Additionally, ``1st galaxies'' can
potentially enable observations of the first stages of galaxy
formation.

Halo clustering can, in principle, be used to constrain both cosmology
and the physics of galaxy formation with extremely high redshift
galaxies ($z\simgt6$) as found, for example, by the Lyman-break
(dropout) technique (\eg \citealt{steidel}) or by the redshifted Lyman
alpha emission line (\eg \citealt{subaru}).  We note that our redshift
10 haloes include the likely mass range of candidate $z\sim10$
Lyman-$\alpha$ emitters found behind a lensing cluster by Stark \etal
(2007).  Because the survey volume of that study is highly uncertain,
so is the abundance, and hence the mass, of dark halo hosts of these
objects.  As future ever more sensitive observations target similar
galaxies, and with the upcoming launch of James Webb Space Telescope
(JWST), their clustering properties should be measurable.  This will
constrain the properties of the haloes in which they live.  It is thus
essential to develop our theoretical understanding of halo and galaxy
clustering in this regime.

In this study, we examine halo bias in the reionizing universe.  In
\S~2, we review briefly theoretical studies of halo bias. In \S~3, we
discuss our suite of simulations and analysis techniques.  We present
measurements of halo bias and its scale-dependence from our
simulations in \S~4.  In \S~5, we analyze the kinematics of haloes and
mass, and address the sensitivity of the pairwise velocity dispersion
to the volume of the simulation.  We conclude with a discussion of how
our results can be used as a basis for using reionizing era galaxies
to explore cosmology and galaxy formation.

\section{halo clustering theory}

Clustering can be quantified by counting the numbers of pairs as a
function of separation relative to that of a random distribution, as
in the two-point correlation function,
\begin{equation}
\xi(r)=N_{pairs}(r)/N_{pairs, random}(r)-1.
\end{equation}
Haloes can be either {\it more} or {\it less} clustered than the matter
fluctuation spectrum, as described by the {\it bias}, which we compute
as
\begin{equation}
b = {\sqrt{\xi_{hh}(r)/\xi_{mm}(r)}}, 
\end{equation}
where $\xi_{hh}$ is
the halo two-point correlation function and $\xi_{mm}$ is the same
for the mass.  The two-point correlation function is directly related
to the power spectrum by the Fourier transform, expressed as:
\begin{equation}
\xi(r) = {1 \over 2\pi^2}\int_{0}^{\infty} P(k){\sin(kr) \over kr}
k^{2} dk.
\end{equation}

On large, linear scales, halo formation modeled by extended Press \&
Schechter formalism (\eg \citealt{bower}; \citealt{lc93}) can be used
to predict halo clustering.  In this model, halo formation occurs when
linear fluctuations within a global Gaussian random density field grow
gravitationally to reach a critical threshold for collapse,
$\delta_c$, computed for spherical overdensities.  \cite{mowhite}
utilized this model to predict halo bias and demonstrated that bias
should be scale-independent on sufficiently large scales.  In this
formalism, the bias is closely related to the mass function, which has
been shown by \cite{st99}, Sheth, Mo \& Tormen (2001; SMT), and
\cite{st02} to be better described by an ellipsoidal halo collapse
model. The ellipsoidal collapse model of SMT results in
large-scale bias consistent with simulations of \eg \cite{st99} and
\cite{colberg}, and given by:
\begin{eqnarray}
b_{SMT} = 1 + {1 \over \sqrt{a}\delta_c} \bigg[ \sqrt{a}(a\nu^2) + \sqrt{a}b(a\nu^2)^{1-c} \\ - {(a\nu^2)^c \over (a\nu^2)^c+b(1-c)(1-c/2)} \bigg], \nonumber
\label{smtbias}
\end{eqnarray}
where $a=0.707$, $b=0.5$, $c=0.6$, $\nu=\delta_c/\sigma(m)$,
$\delta_c=1.686$, and $\sigma(m)$ is the RMS linear overdensity in
top-hat spheres of mass m.  For an infinite volume,
\begin{equation}
\sigma^2(m)  =  {D^{2}(z)\over2\pi^2}\int_0^\infty
k^2P(k)W^2(k;m){\rm d}k,
\label{varinf}
\end{equation}
where $P(k)$ is the linear power spectrum of the density fluctuations
at $z=0$, $W(k;m)$ is the Fourier transform of the real-space top-hat
filter, and ${\it D(z)}$ is the growth factor of linear perturbations
normalised to unity at $z=0$ (\citealt{peebles93}). The above bias
relation is universal in the sense that bias depends only upon
$\nu=\delta_c/\sigma(m_{halo})$, which describes the ``rarity'' of the
density fluctuation from which the halo is formed, independently of
halo mass and redshift.  There is an intrinsic scale-dependence of
halo bias, which can be strong on small-scales, but vanishes on
large-scales.  The scale-dependence of bias is difficult to compute
analytically because the mass fluctuation field becomes nonlinear on
small scales.  It is thus advantageous to employ numerical
simulations.

\section{techniques and analysis}

\subsection{The simulations}

This study consists of a subset of the simulations presented in
\cite{reedmf07}, summarized in Table \ref{simtable}.  These runs
consist of dark matter particles evolved from linear initial
conditions using a modified version of the parallel N-body code
{\small GADGET2} (\citealt {gadget2}; \citealt{mill}).  The wide range
of simulation volumes allows us to model haloes over a large range in
mass, and to determine empirically the effects of finite simulation
volume, discussed below.  Consequently, our suite of simulations has a
large enough effective dynamic range to model haloes covering the bulk
of the mass range able to form stars in the reionizing era.  Haloes are
selected by the friends-of-friends algorithm (\citealt{davis}) where
all particles separated by less than $0.2$ times the mean
inter-particle separation are linked into a common halo.
We have tested that our results are not sensitive to the choice of
halo finder by verifying that halos defined using a spherical overdensity
of 178 times mean, another commonly used halo definition, have clustering
properties similar to our friends-of-friends halos (see \S~4.1 for further
discussion).

\subsection{Finite volumes and their effects on spatial bias}

\begin{figure*}
  \includegraphics[height=.75\textheight]{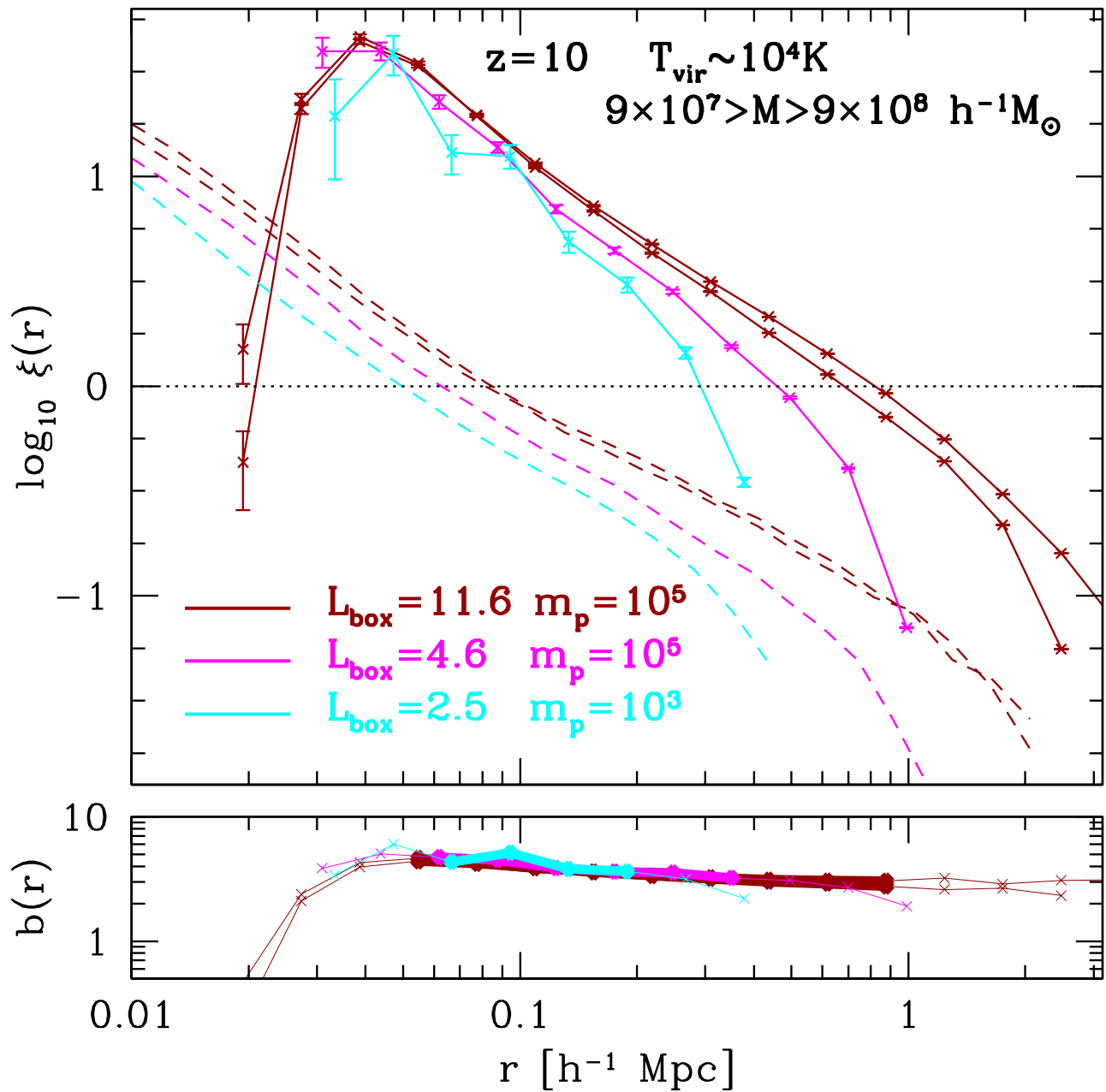}
  \caption{{\it Top panel}: Correlation function for haloes
($\xi_{hh}$, solid) and particles ($\xi_{mm}$, dashed) for all haloes
of mass $\sim10^{8-9} h^{-1} \msun$ extracted from a range of
simulation volumes, at $z=10$.  Here, and in subsequent figures, box
size and particle mass labels have units of $h^{-1}$Mpc and
$h^{-1}\msun$, respectively.  Error bars on $\xi_{hh}$ (solid) are
$1-\sigma$ Poisson from halo pair counts per bin.  {\it Bottom panel:}
Bias, $(\xi_{hh}/\xi_{mm})^{1/2}$, for a range of simulation volumes.
Bold portions denote the most reliable range of scales for the
computed bias, extending to approximately ${\rm 0.1L_{box}}$; the
agreement of different box sizes demonstrates the weakness of finite
volume effects on halo bias.  }
  \label{biasbox}
\end{figure*}

The effects of the finite simulation volume might be expected to alter
the clustering properties of the simulation.  Finite simulation
volumes lack power with wavelengths larger than the box, and suffer
from discreteness of power especially for long wavelength modes (\eg
\citealt{bark}; \citealt{bagray}; \citealt{powerbox}; \citealt{sirko};
\citealt{baglamf}; \citealt{reedmf07}; Luki{\'c} \etal 2007).  
The resulting effects on clustering can be strong for the 
small volumes required to model the low mass haloes relevant
for high redshift star formation.
But fortunately, finite-volume effects are similar for dark
matter and for haloes, as we show in Fig. \ref{biasbox}.
As a result, the finite-volume effects on halo
bias are much weaker than the effects on absolute clustering
properties of mass or haloes.  For this reason, the halo bias measured
within a simulation is reliable over a substantial range of scales
smaller than the length of the box.

Fig. \ref{biasbox} shows an example of the two-point correlation
function, $\xi(r)$, for dark matter and for haloes with virial
temperatures greater than $10^4K$ at $z=10$ (approximately $\sim
10^{8} h^{-1}M_{\odot}$, using the relation
$M_{10^4}=10^8\times[10/(1+z)^{3/2}] h^{-1} \msun$), and spanning a
factor of ten in halo mass.  The bottom panel shows halo bias for
several box sizes, demonstrating that any systematic affects on bias
due to finite-volume are weak provided that the scale sampled is
sufficiently smaller than the box size.  We find that as a
conservative choice, the bias measured from pair separations below
${\rm 0.1L_{box}}$ is adequate for our desired level of accuracy
($\simlt 10\%$) in our simulations.  The bold portion of the bias
curves in Fig. \ref{biasbox} and \ref{biashalos} extend only to this
scale.  Although the scatter in bias is close to $\sim 50\%$ at small
scales and large masses where numbers of pairs are small, we
demonstrate later that we are able to measure bias at large scales to
$\simlt 10\%$, using ${\rm 0.1L_{box}}$ as a maximum pair separation.
Our smallest box size of $1 h^{-1}$Mpc, which might be expected to be
most likely to suffer finite-volume effects, does not have enough
haloes of this mass for us to compute reliably a correlation function;
however, we have confirmed using smaller haloes that the bias found in
the smallest boxes agrees with that in larger boxes.

The decrease in bias at small scales is mainly due to the fact that it
is not possible to have a halo pair separated by less than twice the
halo radius, sometimes referred to as the halo exclusion effect
(\citealt{bensonbias}).  On very small scales, the uncertainty
increases due to small numbers of closely separated pairs, as
reflected by the increased scatter.

We have just shown that halo bias is not affected strongly by box
size, an argument that is strengthened by the agreement among
overlapping boxes in Fig \ref{biashalos}. However, because individual
realizations and different simulation volumes each have a unique form
of $\sigma(m)$, consideration of the universality of halo bias can be
strengthened if one uses the relation between $\sigma$ and $M$
specific to each realization as the effective universal mass variable.
This mitigates the effects of missing large-scale power and reduces
run-to-run scatter.  For a periodic cosmological simulation, the
smoothed rms linear overdensity, $\sigma$, is given by the discrete
analog of Eqn.~\ref{varinf}:
\begin{equation}
\sigma^2(m)  =  {D^2}(z)\sum_{\bf k}|\delta_{\bf k}|^2 W^2(k;m),
\label{varsim}
\end{equation}
where $|\delta_{\bf k}|$ refers to the linear amplitude of the Fourier
modes at $z=0$, and $D$ and $W$ are the same as in Eqn.~\ref{varinf}.
This approach has been successful in the parameterization of the halo mass
function (see \citealt{reedmf07} for detailed discussion).  We discuss
further in the Appendix the importance of considering 
finite volume effects for estimates of large-scale clustering.

\begin{table}
\begin{tabular}{lllll}
\hline\hline
 N$_{\rm runs}$   &  L$_{\rm box}$  & m$_{\rm part}$  & N$_{\rm part}$
& r$_{\rm soft}$ \\ 
    &  $h^{-1}$Mpc & $h^{-1} \msun$ &   & $h^{-1}$kpc \\
\hline
11 & 1.0 & 1.1 $\times$ 10$^{3}$ & 400$^{3}$ & 0.125 \\
1 & 2.5 & 1.1 $\times$ 10$^{3}$ & 1000$^{3}$ & 0.125 \\
1 & 4.64 & 1.1 $\times$ 10$^{5}$ & 400$^{3}$ & 0.58 \\
2 & 11.6 & 1.1 $\times$ 10$^{5}$ & 1000$^{3}$ & 0.58 \\
1 & 20 & 8.7 $\times$ 10$^{6}$ & 400$^{3}$ & 2.5 \\
2 & 50 & 8.7 $\times$ 10$^{6}$ & 1000$^{3}$ & 2.4 \\
\hline

\label{simtable}
\end{tabular}
\caption{
Summary of the simulations presented in this work.  N$_{\rm runs}$
random realisations of cubical volumes of side L$_{\rm box}$ were
simulated.  N$_{\rm part}$ particles of mass m$_{\rm part}$ and
gravitational force softening length r$_{\rm soft}$ were employed.  
}
\end{table}

\section{results: halo bias}
\subsection{Scale-dependence}

\begin{figure*}
  \includegraphics[height=.75\textheight]{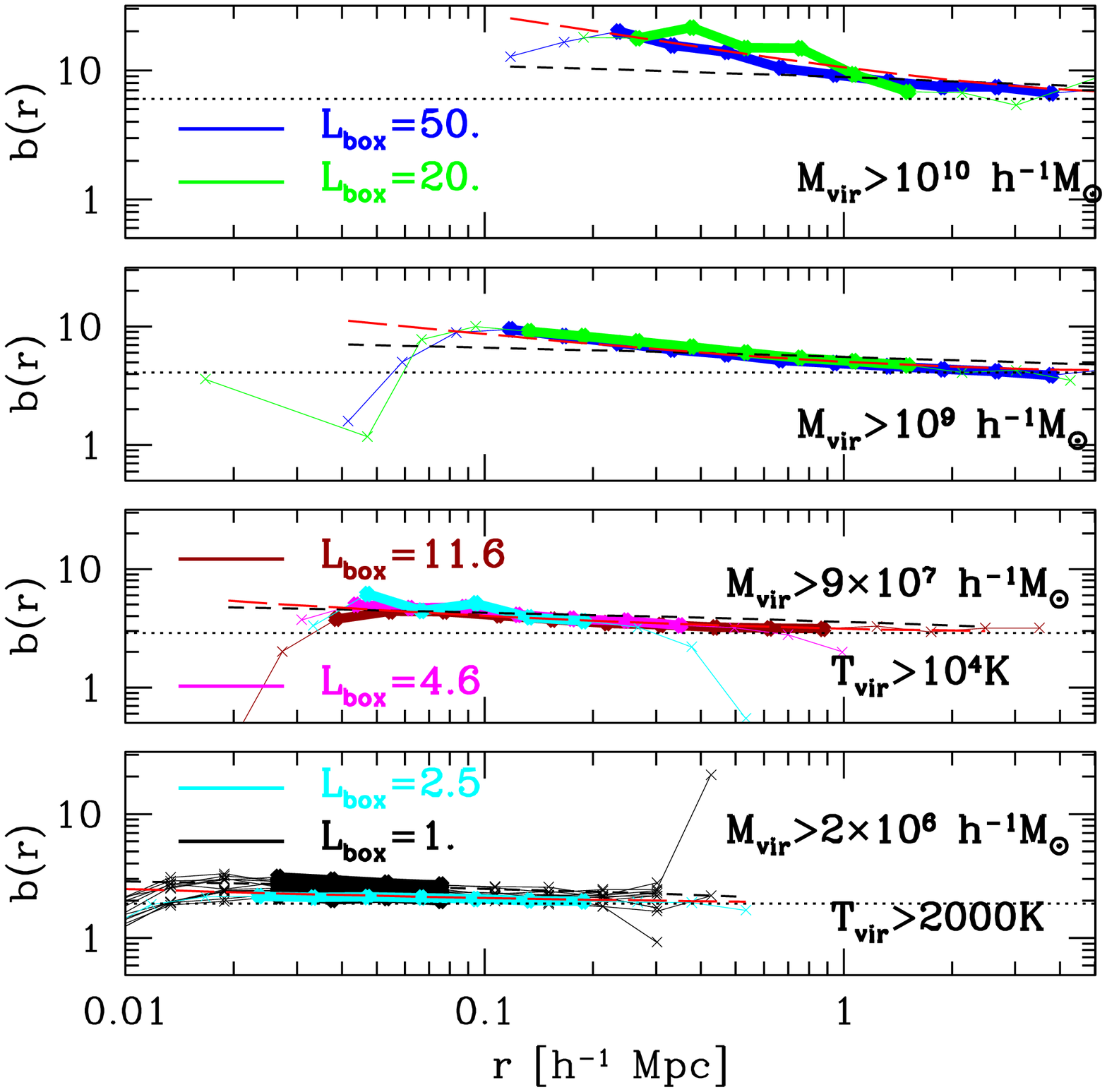}
  \caption{Bias, $(\xi_{hh}/\xi_{mm})^{1/2}$ at $z=10$ for a range of
  halo masses.  The bold portions of the curves denote the most
  reliable range of scales for each box (see text), and long-dashed
  (red) is our fit (neglecting halo exclusion at small scales).  For
  larger haloes, our bias scale-dependence is steeper than
  extrapolations of low redshift fits from Diaferio \etal (2003)
  (short-dashed, black) and others (see text).  Our large-scale bias
  agrees with the prediction by Sheth, Mo, \& Tormen (2001), denoted
  by the horizontal dotted (black) lines.  }
\label{biashalos}
\end{figure*}

We present halo bias as a function of scale for haloes of a range of
masses at redshift 10 in Fig. \ref{biashalos}.  At large scales, the
halo bias approaches the number-weighted large-scale bias predictions
of SMT, given by
\begin{equation}
\label{e.bg}
b_{SMT}(>m_{min}) = n(>m_{min})^{-1}\int_{m_{min}}^\infty \,b_{SMT}(m)\, \frac{dn}{dm}\,dm,
\end{equation}
where $b_{SMT}(m)$ is given in eqn. 4, and we take $dn/dm$ to be the
mass function given by \cite{reedmf07}, Eqn. 11.  We explore further
the large-scale bias in the next subsection.  First, we show the
scale-dependence and compare with fits from clusters in simulations of
much lower redshifts.  The scale-dependence has been parameterized in
terms of the halo pair separation, $r$:
\begin{equation}
b(m,r,z) = b_{ST}(m,z)[1 + b_{ST}(m,z)\sigma(r,z)]^{\alpha},
\end{equation}
with $\alpha=0.15$ by \cite{hamana} and $\alpha=0.35$ by
\cite{diaferio}, where $b_{ST}$ is the bias relation of \cite{st99},
which is similar to $b_{SMT}$.  The steeper scale-dependence of
\cite{diaferio} is consistent with our small haloes, but is not steep
enough to match the most massive haloes in our simulations.  The
\cite{hamana} bias relation (not plotted) is much shallower than that
of \cite{diaferio}, making for a poor match to the scale-dependence in
our data.  The bias fit of \cite{tinkerbias}, which is parameterized
in terms of the non-linear mass correlation function, $\xi(r)$, rather
than the linear variance in spheres, as in the above relation, implies
even weaker scale dependence than that of \cite{hamana} when
extrapolated to our mass and redshift range, and so is not consistent
with our results.

It requires a large extrapolation to apply low redshift fits of
cluster bias to our higher redshifts, lower masses, and much rarer
(larger $\nu$) haloes than previously tested, so it is perhaps not
surprising that we find a steeper scale-dependence for many of our
haloes.  Even so, the reasonable level of agreement for our lower mass
haloes suggests ``partial universality''.  Our haloes at redshift 10 are
well fit by the following relation:
\begin{equation}
b(m,r,z) = b_{SMT}(m,z)[1 + 0.03b_{SMT}(m,z)^{3}\sigma(r,z)^{2}]
\label{biasrfitsmt}
\end{equation}
(ignoring halo exclusion effects at small separations).  We have
checked that our fit remains consistent with our simulation outputs
from redshifts 10 to 30, though the uncertainties and scatter become
greater at high redshifts.  This fit, however, is not intended to be
valid at redshifts lower than we have considered.  In fact, if
extrapolated to the lower redshifts probed by \cite{diaferio}, our fit
has significantly steeper scale-dependence than their simulations,
reflecting non-universality.  We have confirmed the lack of
universality of this parameterization of the scale-dependent bias
using separate simulations of larger volumes evolved to low redshifts.
Though the halo samples considered here were divided into cumulative
samples for illustrative purposes, the above fit agrees with the bias
obtained when samples are restricted to a much narrower mass range of
a factor of two.  For visual clarity, we have chosen to plot in
Fig. \ref{biashalos} the value of $b_{SMT}$ for an infinite volume
rather than the individual finite-volume corrected value for each
simulation.  We demonstrate in the Appendix that the finite-volume
corrected large-scale bias of each realization is a much better
description of the bias measured in simulations, particularly for the
smallest boxes.

A number of factors may affect the universality of the
scale-dependence of halo bias.  The slope of the linear power spectrum
on halo pre-collapse scales is much steeper in our simulations because
the halo masses are smaller relative to that typical of low redshift
simulations.  The steeper power spectrum 
may affect halo clustering (see \eg \citealt{jing})
and cause the apparent non-universality with respect to low redshift
clusters.  Another contributing factor may be that the effective
definition of a halo changes based on its concentration and
environment (Luki{\'c} \etal 2008).  The low concentrations and dominance
of large filamentary structure at high redshift (Gao \etal 2005) may
cause the halo finder to combine or split haloes differently relative
to their Lagrangian (linear) initial density fluctuations.  We have
checked the effect of halo definition by verifying that the bias of
haloes defined using a spherical overdensity of 178 times mean is
consistent, though perhaps slightly higher and steeper, as might be
expected from their lower abundance (\citealt{reedmf07}).  Further
theoretical work regarding halo formation is needed to understand
fully the scale-dependence of halo clustering.  We leave the search
for a more universal parameterization of the scale-dependence of bias
to future studies.

\subsection{Large-scale bias}

\begin{figure}
  \includegraphics[height=.35\textheight]{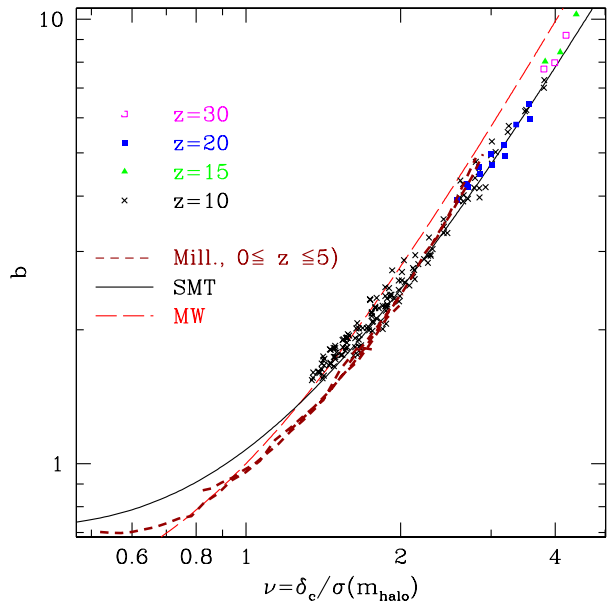}
  \caption{ {\it Large-scale bias}, $[\xi_{hh}/\xi_{mm}]^{1/2}$, shown
 as a function of $\nu=\sigma(m_{halo})/\delta_c$ compared with the
 theoretical predictions of Mo \& White (1996) and Sheth, Mo \& Tormen
 (2001).  Bias is computed in bins of factors of 2 in halo mass, using
 a least-squares fit that assumes the scale-dependence described by
 Eqn. \ref{biasrfitsmt}. Millennium run data are taken from Gao \etal
 (2005).}
\label{biaslarge}
\end{figure}

We compute the large-scale bias for our halo sample in
Fig. \ref{biaslarge}. The SMT bias relation is consistent with our
data at all redshifts while the Mo \& White bias relation is
inconsistent with our results.  By going to very high redshift, we are
able to measure the bias of haloes formed from $\sim4\sigma$
fluctuations, representing a major improvement upon previous bias
measurements at high redshift.  
In particular, the Millennium run bias data (from
\citealt{gaoage}) extend to $3\sigma$ haloes at $z=5$, and
\cite{cohnhiz} compute bias for a narrower mass range at $z=10$.
\cite{angulobias} measure bias for $\sim4.5\sigma$ halos,
but at redshifts of 3
and below.  While we have not explicitly plotted uncertainties, we note
that the scatter among different simulations can be used to estimate
visually an ensemble bootstrap error.  This scatter suggests that the
data are accurate to about 10$\%$ in bias
(ignoring any potential systematics).  Given the scatter within our
data, it is not possible to distinguish between the SMT bias relation
and that of more precise fits such as \cite{seljakbias},
\cite{tinkerbias}, or the bias relation that would be derived using
the mass function fit to these simulations (\citealt{reedmf07}).  

The
agreement with the plotted Millennium data (\citealt{mill}) for our
$2-3\sigma$ haloes confirms universality and suggests that our box-size
correction technique is successful.  The evidence for universality
of halo bias is strengthened by our agreement with \cite{angulobias}
for $2-4\sigma$ haloes, where their results agree with the SMT bias.
However, the source of the $\sim
20\%$ disagreement between our data and the lower redshift bias at
$\nu \simeq 1.5$ from the Millennium run, with which the \cite{angulobias}
bias appears to also agree, is not clear.  Although this
could indicate departure from universality at the $\simeq$20$\%$ level
(in bias), because the differences occur at small masses in the
smallest boxes where the finite-volume corrections are largest (see
Appendix), potential systematics are not easily ruled out.  Our work
verifies that estimates of halo bias derived from extended
Press-Schechter theory and variants are valid for extremely rare
haloes.  Additionally, our comparison with other studies at lower
redshifts significantly extends the confirmed regime of approximate 
universality
(at the $10-20\%$ level) in mass, redshift, and $\nu$ of large-scale
bias.

Some care must be taken when computing bias from haloes in finite
volumes.  Because the scale-dependence of the bias extends to scales
approaching the maximum pair separations where bias measurements
remain robust, a simple averaging of the bias over the reliably 
sampled range in scales would result in a
systematic error (\ie a ``bias bias'').  We thus utilize the scale
dependence of bias presented earlier to fit more reliably the large
scale bias.  The large-scale bias that is plotted in
Fig. \ref{biaslarge} is then computed by performing a least squares
fit using Eqn. \ref{biasrfitsmt} and replacing $b_{SMT}$ with the bias
to be fit, as follows:
\begin{equation}
b(m,r,z) = b_{fit}(m,z)[1 + 0.03b_{fit}(m,z)^{3}\sigma(r,z)^{2}].
\end{equation}
To avoid peculiarities of bias for closely separated pairs where
numbers are small and bias is steepest, bias is then fit over pairs
separated by a range of ${\rm 0.033-0.1L_{box}}$, which also remains
within the range shown previously to be most robust against finite box
effects.

In \cite{cohnhiz}, the large-scale bias was taken to be the bias at
$1.5 h^{-1}$Mpc.  They caution that they may not have captured the
bias at a scale large enough to reflect the asymptotic large-scale
value.  Indeed, their large-scale bias estimates are higher than ours
over the range of overlap, covering 3-3.7$\sigma$ haloes, also at 
$z=10$.  Their bias
estimates are $\simeq$10-20$\%$ higher than the SMT bias relation,
whereas our results agree with the SMT bias.  
These differences are consistent with the
overestimate in bias expected when estimating the large-scale bias
from the value at $1.5 h^{-1}$Mpc (see Fig. \ref{biashalos}).

\section{results: halo pairwise velocity dispersion}

\begin{figure}
  \includegraphics[height=.35\textheight]{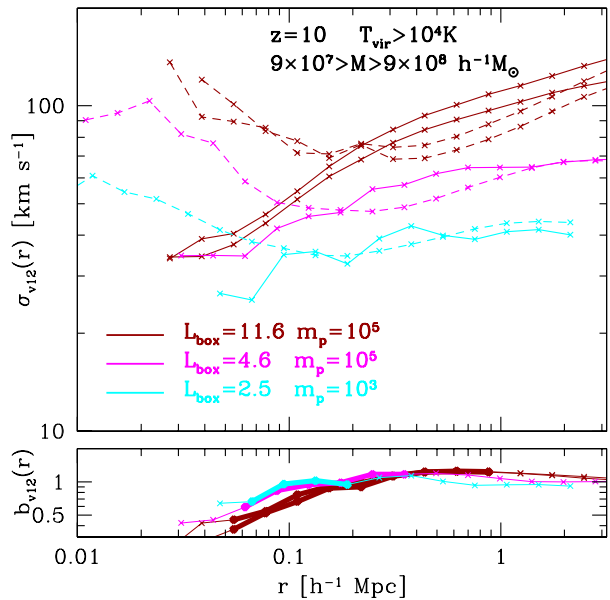}
  \caption{ {\it Top panel}: Pairwise velocity dispersion of halo
pairs ($\sigma_{v12,hh}$(r), solid) and particle pairs
($\sigma_{v12,mm}$(r), dashed) for all haloes with mass $\sim10^{8-9}
h^{-1} \msun$ extracted from a range of simulation volumes, at $z=10$
(the same sample as Fig. \ref{biasbox}).  The finite-volume effect on
pairwise velocities decreases with close pair separation, presumably
because the infall velocity of close pairs is dominated by the total
mass of the pair.  {\it Bottom panel:} Pairwise velocity bias,
$\sigma_{v12,hh}/\sigma_{v12,mm}$ for a range of simulation volumes.
Bold portions denote the most reliable range of scales previously
estimated for {\it spatial} bias (extending to approximately ${\rm
0.1L_{box}}$).  The relative disagreement in pairwise velocities and
bias for simulations of different volume indicate significant finite
volume effects in the kinematics of mass and haloes.}
\label{vbiasbox}
\end{figure}

The kinematics of high redshift galaxies represent an additional probe
of cosmology and galaxy formation, potentially more sensitive than the
spatial correlation function (\eg \citealt{zhaosigmav}).  The pairwise
velocity dispersion, $\sigma_{v12}$(r), has been utilized extensively
in low redshift cosmological surveys (\eg \citealt{davispeebles}).
$\sigma_{v12}$(r) is defined as the one dimensional rms velocity of
pair members in the direction of the line-of-sight connecting the
pair, separated by distance $r$.  From Fig. \ref{vbiasbox} it can be
seen that the pairwise velocity dispersion of both mass and haloes
tends to be suppressed by finite simulation volume.  Particles and
haloes of identical mass in different size volumes have significantly
different velocities; note that the pairwise velocity dispersion has
an inherent mass-dependence, so objects of similar mass must be
considered.  In the case of haloes, some kinematic effects could be due
to the difference in average halo mass of the sample due to the
finite volume suppression of high mass haloes (see \citealt{reedmf07}).
Pairwise dispersions for the $\sim 10^{4}K$ halo sample, (covering
$\sim 10^{8-9}h^{-1} \msun$) at separations of $0.2 h^{-1}$Mpc
(comoving) range from $\simeq75 {\rm km~s^{-1}}$ in the $12 h^{-1}$Mpc
volume to $\simeq50 {\rm km~s^{-1}}$ in the $5 h^{-1}$Mpc volume,
dropping to $\simeq30 {\rm km~s^{-1}}$ in the $2.5 h^{-1}$Mpc volume.
The finite-volume effect on pairwise velocities decreases with close
pair separation, presumably because the infall velocity of close pairs
is dominated by the total mass of the pair.

\begin{figure}
\includegraphics[height=.35\textheight]{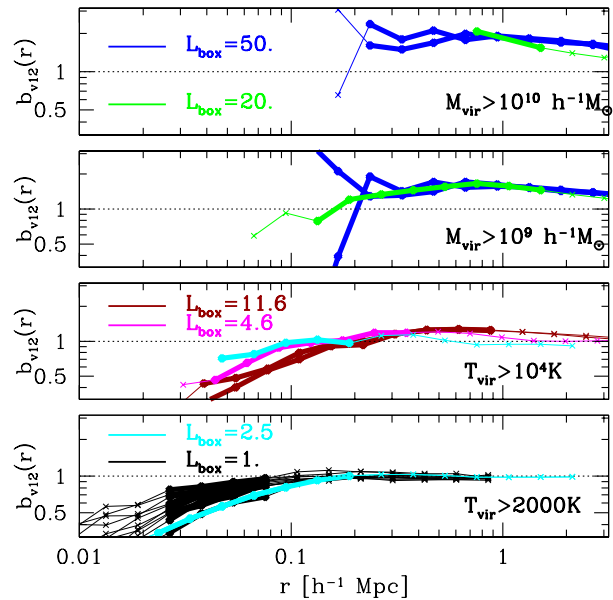}
\caption{Pairwise velocity bias, $\sigma_{v12,hh}/\sigma_{v12,mm}$, at
$z=10$ for the same halo samples as in Fig. \ref{biashalos}.  The bold
portions of the curves denote the range of pair separation where
spatial bias was found to be relatively robust.  In contrast to
spatial bias, however, kinematic bias is affected significantly by
finite simulation volume because particle and halo velocities are
suppressed at different levels.}
\label{vbiashalos}
\end{figure}

Particle and halo kinematics are affected at different levels by
finite volumes, altering the pairwise velocity dispersion bias,
$b_{v12}$, defined here as the ratio of $\sigma_{v12}$(r) for haloes
compared to particles ($\sigma_{v12,hh}/\sigma_{v12,mm}$). Velocity
bias is shown in Fig \ref{vbiashalos} for the same haloes shown in the
clustering bias plot (Fig \ref{biashalos}).  Run-to-run scatter in
velocity bias becomes increasingly large on small scales.  This is in
contrast to the halo spatial bias, which is largely free of finite
volume effects on scales below ${\rm \sim0.1L_{box}}$, despite
similarly strong effects individually on particle and halo clustering
strength.  We leave further development of techniques for correcting
for effects of finite simulation volume on kinematic effects to future
works.

\section{discussion and conclusions}

We have presented results for the clustering of haloes spanning the
mass range likely to host stars and galaxies at redshift ten.
Ultimately, the clustering properties of high redshift galaxies may be
used as a probe of cosmology and also as a probe of the physics of
high redshift galaxy formation.  Though we focus our discussion on the
reionizing era, clustering properties are potentially a valuable tool
at any redshift, particularly for rare haloes where the bias is high
with strong scale-dependence, and the sensitivity to cosmological
parameters and halo mass is strong

\begin{figure}
  \includegraphics[height=.35\textheight]{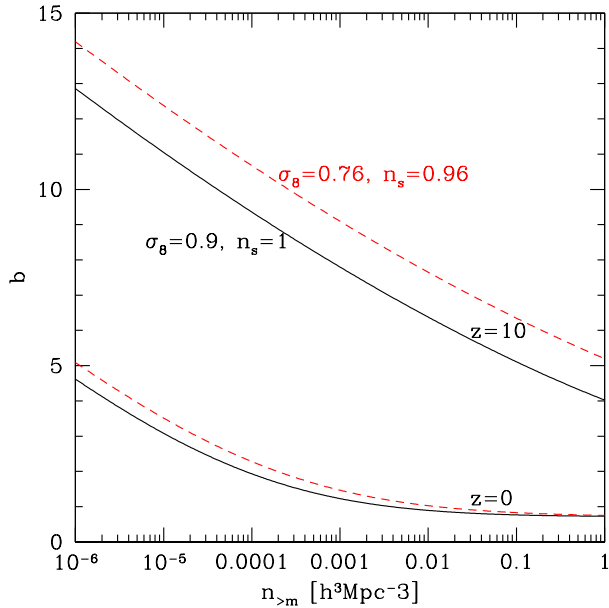}
  \caption{Bias, $[\xi_{hh}/\xi_{mm}]^{1/2}$ at $z=10$ as a
  function of cumulative halo number density for 2 cosmologies with
  different values of $\sigma_8$ and $n_s$.  The bias relation used is
  the Sheth, Mo \& Tormen (2001) fit, applicable to large scales (see
  Fig. \ref{biashalos}).  Halo number density is a fit to simulations
  from Reed \etal (2007), Eqn. 11.  Halo bias at $z=10$ is generally
  more sensitive to cosmology than at low redshifts, and remains so
  over a wide range in halo abundance (or halo mass).}
\label{biasvsn}
\end{figure}

As an example, it is worth considering the simplest case where there
is a one-to-one relation between halo mass and galaxy luminosity, \ie
if $L_{halo}(m,z)$ has no scatter, and exactly one galaxy occupies
each halo.  Here, a measurement of the luminosity function would allow
one to halo derive the halo mass for an assumed cosmology directly
from the mass function.  The inferred mass then determines the halo
bias which, in turn, can be used to constrain cosmological parameters,
as illustrated in Fig. \ref{biasvsn}.  
Note that determining the galaxy bias from observations requires a
simultaneous measurement of both matter clustering and galaxy
clustering, a difficulty which we later discuss.
Since all potentially
observable objects at extremely high redshifts live in haloes much
larger than $M_*$, the cosmology dependence of bias remains large for
all observationally attainable halo abundances.  In contrast, 
at low redshifts, the
cosmology dependence of bias grows weak for small abundances.  For the
most abundant haloes, a bias measurement of $20\%$ accuracy could
distinguish between a low $\sigma_8$ of $0.76$ and a high $\sigma_8$
of $0.9$ if measured from haloes of abundance $\sim 1h^3$Mpc$^{-3}$
($\sim10^{8-9} h^{-1}\msun$).  Although neither of these
normalizations are favored by current CMB data, they are in fact
within $\sim (2-3) \sigma$ of the recent WMAP 5 year normalization
(\citealt{komatsu}).  Moreover, the point remains that bias is
generally more sensitive to cosmology at high redshifts, and at a wide
range of halo abundances and mass.

The bias-abundance relation may remain largely unaltered even if IGM
absorption is significant.  \cite{ilievclust}, for the case of high
redshift Ly-$\alpha$ sources, showed that dimming due to IGM
absorption does not affect significantly the relation between observed
number density and source clustering.  
Despite the likelihood of
increasing galaxy clustering at fixed apparent luminosity
(\citealt{mcquinn}) by altering the apparent luminosity-mass relation,
clustering at fixed number density is not affected strongly by IGM
absorption.
This means that the high
optical depth of the high redshift IGM
should not diminish the potential of using reionizing era galaxies for
cosmology.

The discussion so far is intended to highlight only the {\it
potential} for use of high redshift galaxy clustering measurements to
constrain cosmology.  In practice, to determine halo bias, one must
measure not only the strength of galaxy clustering, but also the
strength of matter clustering.  There is the possibility of measuring
matter clustering at these redshifts using 21cm tomography, although
it will be a challenge because of complex astrophysical effects
including sources of Lyman-$\alpha$ and ionizing photons (see \eg
\citealt{bl21cm}; \citealt{furlanetto}; \citealt{shaw}).  For this
reason, galaxy clustering measurements may have more value as an
indicator of the mass of the halos in which the galaxies reside, once
one has determined the cosmological parameters by other means.  The
steep mass-dependence of both the halo number density and the halo
bias at these high redshifts increases their power as a means of
linking galaxy properties to halo mass.

If there is scatter in the $L_{halo}(m)$ relation, the relation
between abundance and mass no longer holds (\eg \citealt{whiteqso}).
However, provided that on large-scales the scatter is random, there
still remains a direct one to one relation between large-scale bias
and halo mass for a given cosmology via the ``universal'' quantity
$\nu$.  This suggests the potential to constrain the mass of the dark
halo hosts of reionizing galaxies by measuring the clustering
strength.  At slightly lower redshifts of $z\sim4-5$, clustering
measurements have been made of Ly-$\alpha$-emitters and Lyman-break
galaxies, allowing estimates of the mass of their host haloes (see \eg
\citealt{hamanahof}; \citealt{kashikawa}; \citealt{hamana06};
\citealt{lee}; \citealt{conroyhiz}; \citealt{kovac}).

Once the large-scale bias and cosmology are determined, the
scale-dependence of halo bias, to the degree that it is ``universal'',
is also determined.  Comparison of the observed scale-dependence of
galaxy clustering to that of haloes with the same large-scale bias 
thus provides a
valuable indicator of local environmental effects on galaxy formation.
The strong clustering and steep scale-dependence of the clustering for
high redshift haloes may cause feedback from neighbouring galaxies to
affect star formation during these early times.

Many possible complications to this simple picture will need to be
considered.  For example, the large scales of reionization bubbles
(\eg \citealt{ilievreion}; \citealt{zahnreion}) may correlate with
galaxy clustering on large scales (\eg \citealt{wyitheloeb}), which
leads to the possibility of reionization affecting galaxy formation
and galaxy clustering on large scales (\eg \citealt{gnedin}).  Or,
multiple galaxies may occupy massive haloes, which creates a degeneracy
in galaxy abundance between satellite numbers and halo mass.  Similarly,
there exists a degeneracy between halo mass and satellite numbers
in the scale-dependence of clustering on small scales.  These
degeneracies are difficult to overcome observationally because of the
need for clustering measurements on very small angular scales (see \eg
\citealt{bullockocc}).  However, one can use theoretical arguments,
semi-analytic modelling and/or smaller volume simulations utilizing
hydrodynamics and other detailed physics to infer basic galaxy
properties such as the numbers, luminosities, and environmental
influences (\eg \citealt{mcquinn}; \citealt{mesinger};
\citealt{ilievreion}; \citealt{ricotti1st}).  Progress may then be
made by combining the results of such detailed models of galaxy
formation with determinations of halo clustering from dark matter
simulations in order to produce detailed predictions for observable
clustering properties.  This will help to break the degeneracies
between galaxy ``feedback'' and halo clustering on small scales, as
well as determine whether haloes tend to host multiple luminous
galaxies.  As ever deeper observations are made, such studies will
allow us to probe galaxy formation at its earliest stages and to
explore cosmology at the high redshift frontier.

We close with a brief discussion of the (perhaps surprisingly) large
suppression to the halo pairwise velocity dispersion by finite
simulation volumes.  Pairwise velocity dispersions are reduced by up
to and beyond $50\%$ in our simulations, even on scales where halo
spatial bias is well behaved.  The general suppression of halo and
mass velocities may have significant effects on hydrodynamic
simulations where small volumes are often required because of
computational expense.  For example, volumes of $\simlt 1$Mpc are
commonly employed in simulations that model the cooling of gas within
high redshift haloes to form the ``first stars'' (\eg Abel, Bryan, \&
Norman 2000,2002; Bromm, Coppi, \& Larson 1999,2002;
\citealt{yoshida1st}; \citealt{osheanorman}).  This suggests the
possibility that gas heating due to shocks of infalling material of
merging haloes could be inhibited due to the suppression of pairwise
velocities.  Such effects have the potential to alter gas cooling and
the ionization level of halo gas, which could affect star formation.
However, because the velocity suppression diminishes for close pairs,
it is not clear whether the kinematic suppression is actually a major
problem for gas dynamics or star formation within simulations of the
high redshift universe.  Any such problems could be minimized by the
``zoom'' resimulation technique, consisting of a high resolution
subvolume embedded within a larger and lower resolution cosmological
volume, as in \eg \cite{gao1st}.  Further investigation of
finite-volume effects on kinematics and any associated effects on gas
properties is warranted.

\section*{acknowledgments}

DR is a post-doc LANL, and is supported by the DOE through the
IGPP, the LDRD-DR and the
LDRD-ER programs at LANL.  CSF acknowledges a Wolfson Royal Society
research merit award.  We thank Salman Habib, Katrin Heitmann, Zarija
Luki{\'c}, Martin White, Joanne Cohn, Jeremy Tinker, and Brian O'Shea
for helpful discussions.  We thank Volker Springel for use of an
enhanced version of {\small GADGET-2}.  We thank Liang Gao for making
his Millennium simulation results available to us.  We are grateful
for the suggestions of the anonymous referee.

{}

\section{appendix}

\begin{figure}
  \includegraphics[height=.35\textheight]{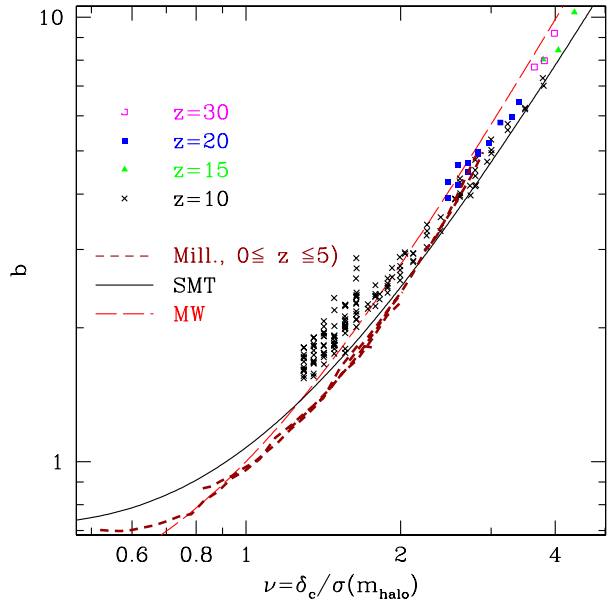}
  \caption{ This figure uses the sigma(m) computed for an infinite
  volume (Eqn. \ref{varinf}) instead of that computed taking into
  account the power present in each realization (Eqn. \ref{varsim}).
  Otherwise this figure is identical to Fig. \ref{biaslarge} (including
  the Millennium run results from Gao \etal 2005).  The
  larger run-to-run scatter in bias and the poorer fit to the SMT
  relation demonstrate the usefulness of correcting for reduced power
  and discrete modes in finite volumes.  }
\label{biastheopow}
\end{figure}

Here we demonstrate the importance of taking into account the finite
simulation volume when considering large-scale halo bias.  By
computing $\sigma^2(m)$ from Eqn. \ref{varsim} instead of
Eqn. \ref{varinf}, the mean power at each discrete wavenumber present
in the simulation is taken into account.  This has the effect of
shifting $\sigma(m_{halo})$, usually toward smaller values (larger
values of $\nu$).  Fig. \ref{biastheopow}, which uses
Eqn. \ref{varinf} to compute $\sigma(m_{halo})$, thereby ignoring the
finite volumes, shows that two large effects are present if finite
simulation volume is not treated appropriately.  First, there is
significant run-to-run scatter in bias, approaching $\sim 50\%$ for
the smallest volume, to be compared with the $\sim 10\%$ scatter in
Fig. \ref{biaslarge} where Eqn. \ref{varsim} is used to compute
$\sigma(m_{halo})$.  Second, the inferred simulation bias is
significantly higher when finite volume is ignored, especially for the
smallest boxes, and the larger haloes within them.  This approach of
finite-volume compensation does not correct for discreteness of
phases, nor does it provide a correction for the deviations as a
result of non-linear coupling to only a small number of modes at large
scales (\citealt{takahashi}).  However, the improvement in run-to-run
scatter, the better fit to the SMT bias relation, and the better agreement
with the Millennium run results, all provide
some level of assurance that this approach of correcting for finite
volume is justified.  Similar reductions in run-to-run scatter and 
improved universality of the halo mass function were
demonstrated in \cite{reedmf07}, strengthening the importance and
validity of correcting for finite simulation volume in cosmological
simulations.  These corrections allow one to simulate smaller haloes,
and hence achieve a larger effective dynamic range, while maintaining
their value as cosmological tools.

\label{lastpage}

\end{document}